\documentclass{article}

\usepackage{times}
\usepackage{amssymb,amsmath,amsthm}
\usepackage{mathrsfs}
\usepackage{epsfig}
\usepackage{color}

\newcommand\ie{{\em i.e.}~}
\newcommand\eg{{\em e.g.}~}

\def\B{\mathscr B}
\def\C{\mathbb C}
\def\D{\mathscr D}
\def\E{\mathcal E}
\def\F{\mathscr F}

\def\H{\mathcal H}

\def\K{\mathcal K}

\def\R{\mathbb R}
\def\S{\mathscr S}

\def\12{{\textstyle\frac12}}
\def\<{\left\langle}
\def\>{\right\rangle}
\def\({\left(}
\def\){\right)}
\def\[{\left[}
\def\]{\right]}
\def\dom{\mathcal D}
\def\lone{\mathsf{L}^{\:\!\!1}}

\def\ltwo{\mathsf{L}^{\:\!\!2}}
\def\linf{\mathsf{L}^{\:\!\!\infty}}

\def\e{\mathop{\mathrm{e}}\nolimits}
\def\d{\mathrm{d}}

\def\im{\mathop{\mathsf{Im}}\nolimits}

\def\sgn{\mathop{\mathrm{sgn}}\nolimits}

\def\slim{\mathop{\hbox{\rm s-}\lim}\nolimits}

\newtheorem{Theorem}{Theorem}[section]
\newtheorem{Remark}[Theorem]{Remark}
\newtheorem{Lemma}[Theorem]{Lemma}
\newtheorem{Assumption}[Theorem]{Assumption}

\newtheorem{Proposition}[Theorem]{Proposition}

\newtheorem{Example}[Theorem]{Example}

\begin{document}


\title{{\Large\textbf{Time delay for dispersive systems in quantum\\
scattering theory. I. The Friedrichs model}}}

\author{Rafael Tiedra de Aldecoa}
\date{\small
\begin{quote}
\emph{
\begin{itemize}
\item[] CNRS (UMR 8088) and Department of Mathematics, University of Cergy-Pontoise,
2 avenue Adolphe Chauvin, 95302 Cergy-Pontoise Cedex, France
\item[] \emph{E-mail:} rafael.tiedra@u-cergy.fr
\end{itemize}
}
\end{quote}
}

\maketitle


\begin{abstract}
We present a method for proving the existence of time delay (defined in terms of sojourn times)
as well as its identity with Eisenbud-Wigner time delay in the case of the Friedrichs model. We
show that this method applies to scattering by finite rank potentials.
\end{abstract}

\section{Introduction}\label{Introduction}

One can find a large literature on the identity of Eisenbud-Wigner time delay and time delay
in quantum scattering defined in terms of sojourn times (see \cite{AC87,AS06,BGG83,dCN02,GS80,JSM72,Jen81,Mar76,Mar81,MSA92,Nak87,Nar84,Rob94,RW89,Wan88}
and references therein). However, most of the papers treat scattering processes where the
free dynamics is given by a Schr\"odinger operator. The mathematical articles where different
scattering processes are considered (such as \cite{JSM72,Mar76,Mar81,Rob94}) only furnish
explicit applications in the Schr\"odinger case. The purpose of the present paper and the
forthcoming work \cite{Tie09} is to fill in this gap
by proving the existence of time delay and its identity with Eisenbud-Wigner time delay for a general
class of dispersive quantum systems. Using a symmetrization argument introduced in
\cite{BO79,Mar81,Smi60} for $N$-body scattering, and rigorously applied in \cite{AJ07,GT07,Mar75,Tie08,Tie06},
we shall treat any two-body scattering process with free dynamics given by a regular enough
pseudodifferential, or multiplication, operator. In this paper we restrict ourselves to the simple, but instructive, quantum model introduced by Friedrichs \cite{Fri38}. The general case will be considered
elsewhere \cite{Tie09}.

Let $H_0:=Q$ be the position operator in the Hilbert space $\H:=\ltwo(\R)$, endowed with the norm
$\|\cdot\|$ and the scalar product $\<\cdot,\cdot\>$. Let $H$ be a selfadjoint
perturbation of $H_0$ such that the wave operators $W_\pm:=\slim_{t\to\pm\infty}\e^{itH}\e^{-itH_0}$ exist and are complete (so that the scattering operator $S:=W_+^*W_-$ is unitary).
Take a (localization) function $f\in\linf(\R;\C)$ decaying sufficiently fast at
infinity. Then we define for some states $\varphi\in\H$ and $r>0$ the quantities
\begin{align*}
T^0_r(\varphi)&:=\int_\R\d t\<\e^{-itH_0}\varphi,f(P/r)\e^{-itH_0}\varphi\>,\\
T_r(\varphi)&:=\int_\R\d t\<\e^{-itH}W_-\varphi,f(P/r)\e^{-itH}W_-\varphi\>,\\
\tau_r^{\rm in}(\varphi)&:=T_r(\varphi)-T^0_r(\varphi),\\
\tau_r(\varphi)&:=T_r(\varphi)-\12\[T^0_r(\varphi)+T^0_r(S\varphi)\],
\end{align*}
where $P$ is the momentum operator in $\H$. If $\|\varphi\|=1$ and $f$ is the characteristic function
$\chi_J$ for a bounded set $J\subset\R$, then these numbers admit a simple interpretation. The
operator $f(P/r)\equiv\chi_{rJ}(P)$ is the orthogonal projection onto the set of states with momentum localised
in $rJ:=\{x\in\R\mid x/r\in J\}$. Therefore $T^0_r(\varphi)$ is the time spent by the freely evolving
state $\e^{-itH_0}\varphi$ in the subset $\chi_{rJ}(P)\H$ of $\H$ (\ie the time during which
$\e^{-itH_0}\varphi$ has momentum in $rJ$). Similarly $T_r(\varphi)$ is the time spent by
the associated scattering state $\e^{-itH}W_-\varphi$ in $\chi_{rJ}(P)\H$. Then
$\tau_r^{\rm in}(\varphi)$ is the time delay in $\chi_{rJ}(P)\H$ of the scattering process with incoming
state $\varphi$, and $\tau_r(\varphi)$ is the corresponding symmetrized time delay.
One can give an equivalent interpretation, with momenta replaced by positions, by using the Fourier
transformation.

In this paper we study the existence of $\tau_r^{\rm in}(\varphi)$ and $\tau_r(\varphi)$ as
$r\to\infty$. Under general assumptions on $f$, $H$, and $\varphi$ we show in Lemma
\ref{lemma_free}.(b) that
$$
\lim_{r\to\infty}\tau_r^{\rm in}(\varphi)=\lim_{r\to\infty}\tau_r(\varphi)
$$
whenever one of the two limits exists. In Theorem \ref{abs_delay} and Remark \ref{Remark_Eisenbud} we prove the
Eisenbud-Wigner formula for the Friedrichs model in an abstract setting. For general $f$, $H$, and
$\varphi$, we show that
\begin{equation}\label{EWintro}
\lim_{r\to\infty}\tau^{\rm in}_r(\varphi)=-i\int_\R\d x\,|\varphi(x)|^2\overline{S(x)}S'(x)
\end{equation}
if the scattering matrix $x\mapsto S(x)$ is continuously differentiable on the support of $\varphi$.
Some comments on the relation between Equation \eqref{EWintro} and the Birman-Krein formula are
given in Remark \ref{RemarkBirman}. In Section \ref{finite_rank} we verify the assumptions of
Theorem \ref{abs_delay} and Remark \ref{Remark_Eisenbud} when $H$ is a regular enough finite rank perturbation of $H_0$. The main difficulty consists in showing (as in the Schr\"odinger case \cite{ACS87,JN92})
that the scattering operator maps some dense set $\E\subset\H$ into itself.
Essentially this reduces to proving that the scattering matrix is sufficiently differentiable on
$\R\setminus\sigma_{\rm pp}(H)$, which is achieved by proving a stationary formula for $S(x)$
and by using higher order commutators methods (see Lemmas \ref{rhube}-\ref{S_differentiable}).
All these results are collected in Theorem \ref{final_one}, where Equation \eqref{EWintro} is proved
for finite rank perturbations. Some properties of a restriction operator \cite[Chap. 2.4]{Kur78}
are recalled in the appendix.

We emphasize that our approach relies crucially on the proof of the propagation formula
\begin{equation}\label{not_ugly}
\lim_{r\to\infty}\int_0^\infty\d t\,\big\langle\varphi,\big[\e^{itH_0}f(P/r)\e^{-itH_0}
-\e^{-itH_0}f(P/r)\e^{itH_0}\big]\varphi\big\rangle=2\big\langle\varphi,P\varphi\big\rangle,
\end{equation}
which relates the time evolution of the localization operator $f(P/r)$ to the energy derivative $iP\equiv\frac\d{\d H_0}$  (see Proposition \ref{equal_P}). It allows us to establish Equation \eqref{EWintro} for a general class of localization functions $f$ not considered before (see
Assumption \eqref{assumption_f}). In \cite{Tie09} we will generalise Equation \eqref{not_ugly}
to the case of pseudodifferential operators $H_0$.

We finally mention the paper \cite{AABCDF06} for a related work on sojourn time for the Friedrichs
model.

\section{Propagation formula for $H_0=Q$}\label{integrated_propagation}

We give here the proof of Equation \eqref{not_ugly} under appropriate assumptions on the
localization $f$ and the vector $\varphi$.

\begin{Assumption}\label{assumption_f}
The function $f\in\linf(\R;\C)$ satisfies the following conditions:
\begin{enumerate}
\item[(i)] $f(x)=f(|x|)$ for a.e. $x\in\R$.
\item[(ii)] There exists $\rho>1$ such that $|f(x)|\le{\rm Const.}\<x\>^{-\rho}$ for a.e. $x\in\R$.
\item[(iii)] There exists $\delta>0$ such that $f=1$ on $(-\delta,\delta)$.
\end{enumerate}
\end{Assumption}

It is clear that $\slim_{r\to\infty}f(P/r)=1$ if $f$ satisfies Assumption \ref{assumption_f}. The
typical example of function $f$ one should keep in mind is the following.

\begin{Example}\label{projection}
Let $f=\chi_J$, where $J\subset\R$ is bounded, symmmetric (\ie $J=-J$), and contains an interval
$(-\delta,\delta)$ for some $\delta>0$. Then $f$ satisfies Assumption \ref{assumption_f}, and
$f(P/r)$ is the orthogonal projection onto the set of states with momentum localised in $rJ$.
\end{Example}

For each $s,t\in\R$, we denote by $\H^s_t$ the usual weighted Sobolev space over $\R$, \ie the
completion of the Schwartz space $\S$ on $\R$ for the norm
$\|\varphi\|_{\H^s_t}:=\|\<P\>^s\<Q\>^t\varphi\|$, $\<\;\!\cdot\;\!\>:=(1+|\cdot|^2)^{1/2}$.
We also set $\H^s:=\H^s_0$ and $\H_t:=\H_t^0$.

\begin{Proposition}\label{equal_P}
Let $f$ satisfy Assumption \ref{assumption_f}. Then we have for each $\varphi\in\H^s$, $s>1$,
$$
\lim_{r\to\infty}\int_0^\infty\d t\,\big\langle\varphi,\big[\e^{itQ}f(P/r)\e^{-itQ}
-\e^{-itQ}f(P/r)\e^{itQ}\big]\varphi\big\rangle=2\big\langle\varphi,P\varphi\big\rangle.
$$
\end{Proposition}

\begin{proof}
Let $r>0$. Using the formula
\begin{equation}
\textstyle\e^{itQ}f(P/r)\e^{-itQ}=f\big(\frac{P-t}r\big),\quad t\in\R,\label{evolution}
\end{equation}
one gets
\begin{align*}
I_r&:=\int_0^\infty\d t\,\big\langle\varphi,\big[\e^{itQ}f(P/r)\e^{-itQ}
-\e^{-itQ}f(P/r)\e^{itQ}\big]\varphi\big\rangle\\
&=\int_0^\infty\d t\int_\R\d k\,|(\F\varphi)(k)|^2
\textstyle\big[f\big(\frac{k-t}r\big)-f\big(\frac{k+t}r\big)\big],
\end{align*}
where $\F$ stands for the Fourier transformation. Due to Assumption \ref{assumption_f}.(ii) one
can apply Fubini's theorem to interchange the order of integration. This together with Assumption
\ref{assumption_f}.(i) implies that
$$
I_r=2r\int_\R\d k\,|(\F\varphi)(k)|^2\sgn(k)\int_0^{|k|/r}\d t\,f(t).
$$
By Assumption \ref{assumption_f}.(iii) there exists $\delta>0$ such that $f(t)=\chi_{[0,\delta)}(t)+\chi_{[\delta,\infty)}(t)f(t)$ for each $t\ge0$. Thus
$I_r=I_r^{(1)}+I_r^{(2)}$ with
$$
I_r^{(1)}:=2r\int_\R\d k\,|(\F\varphi)(k)|^2\sgn(k)\int_0^{|k|/r}\d t\,\chi_{[0,\delta)}(t),
$$
and
$$
I_r^{(2)}:=2r\int_\R\d k\,|(\F\varphi)(k)|^2\sgn(k)\int_0^{|k|/r}\d t\,
\chi_{[\delta,\infty)}(t)f(t).
$$
Since $\varphi\in\H^s$ for some $s>1$, one has
\begin{align}
\big|I_r^{(2)}\big|&\leq{\rm Const.}\,r\int_\R\d k\,|(\F\varphi)(k)|^2
\chi_{[\delta r,\infty)}(|k|)|k|/r\nonumber\\
&\le{\rm Const.}\;\!\big\|\<P\>^{1/2}\chi_{[\delta r,\infty)}(|P|)\varphi\big\|^2\nonumber\\
&\le{\rm Const.}\;\!\big\|\<P\>^{1/2-s}\chi_{[\delta r,\infty)}(|P|)\big\|^2
\big\|\<P\>^s\varphi\big\|^2\nonumber\\
&\le{\rm Const.}\<\delta r\>^{1-2s}.\label{vers0}
\end{align}
Thus $\lim_{r\to\infty}I_r=\lim_{r\to\infty}I_r^{(1)}$. Since
$$
\int_0^{|k|/r}\d t\,\chi_{[0,\delta)}(t)
=\delta\chi_{[\delta r,\infty)}(|k|)+\frac{|k|}r\,\chi_{[0,\delta r)}(|k|),
$$
one has
\begin{align*}
I_r^{(1)}
=2\delta r\int_\R\d k\,|(\F\varphi)(k)|^2\sgn(k)\chi_{[\delta r,\infty)}(|k|)
+2\int_\R\d k\,k|(\F\varphi)(k)|^2\chi_{[0,\delta r)}(|k|).
\end{align*}
But calculations similar to \eqref{vers0} show that
$$
\textstyle\big|2\delta r\int_\R\d k\,|(\F\varphi)(k)|^2\sgn(k)\chi_{[\delta r,\infty)}(|k|)\big|
\le{\rm Const.}\,r^{1-s}.
$$
It follows that
$$
\lim_{r\to\infty}I_r=\lim_{r\to\infty}I_r^{(1)}
=\lim_{r\to\infty}2\int_\R\d k\,k|(\F\varphi)(k)|^2\chi_{[0,\delta r)}(|k|)
=2\<\varphi,P\varphi\>,
$$
which proves the claim.
\end{proof}

\section{Time delay}\label{time_delay}

In this section we prove the existence of time delay for the Friedrichs model in $\H$ with free
Hamiltonian $H_0=Q$ and full Hamiltonian $H$. The full Hamiltonian $H$ can be any selfadjoint
operator in $\H$ satisfying the following condition (we write $\B(\H_1,\H_2)$ for the set of
bounded operators from a Hilbert space $\H_1$ to a Hilbert space $\H_2$).

\begin{Assumption}\label{wave}
The wave operators $W_\pm$ exist and are complete, and any operator $T\in\B(\H^{-s},\H)$,
with $s>1/2$, is locally $H$-smooth on $\R\setminus\sigma_{\rm pp}(H)$.
\end{Assumption}

For each $s\ge0$ we introduce the set
$$
\D_s:=\left\{\varphi\in\H^s\mid\chi_J(Q)\varphi=\varphi
\textrm{ for some compact set }J\subset\R\setminus\sigma_{\rm pp}(H)\right\}.
$$
It is clear that $\D_s$ is dense in $\H$ if $\sigma_{\rm pp}(H)$ is of Lebesgue measure $0$
and that $\D_{s_1}\subset\D_{s_2}$ if $s_1\ge s_2$. Furthermore if $\varphi\in\D_0$, then
$T_r(\varphi)$ is finite for each $r>0$ due to Assumption \ref{wave}.

For each $r>0$, we define
\begin{align}
\tau^{\rm free}_r(\varphi)&:=\12\int_0^\infty\d t\<S\varphi,\big[\e^{itH_0}f(P/r)\e^{-itH_0}
-\e^{-itH_0}f(P/r)\e^{itH_0}\big]S\varphi\>\label{tau_free}\\
&\quad-\12\int_0^\infty\d t\<\varphi,\big[\e^{itH_0}f(P/r)\e^{-itH_0}
-\e^{-itH_0}f(P/r)\e^{itH_0}\big]\varphi\>.\nonumber
\end{align}
The number $\tau^{\rm free}_r(\varphi)$ (which has the dimension of a time if $f(P/r)$ is an
orthogonal projection) is finite for all $\varphi\in\H$. We refer the reader to
\cite[Eq. (93) \& (96)]{AJ07}, \cite[Eq. (4.1)]{GT07}, and \cite[Sec. 2.1]{Tie06} for similar
definitions when $H_0$ is the free Schr\"odinger operator and $f(P/r)$ is an orthogonal projection.
The usual definition can be found in \cite[Eq. (3)]{AC87}, \cite[Eq. (6.2)]{Jen81}, and
\cite[Eq. (5)]{Mar76}.

\begin{Lemma}\label{lemma_free}
Let $f$ satisfy Assumption \ref{assumption_f} and suppose that Assumption \ref{wave} holds. Then
\begin{enumerate}
\item[(a)] For each $r>0$ and $\varphi\in\H$ one has the identities
\begin{equation}\label{identities}
T^0_r(\varphi)=r\|\varphi\|^2\int_\R\d x\,f(x),\quad
T^0_r(\varphi)=T^0_r(S\varphi),\quad
\tau_r^{\rm in}(\varphi)=\tau_r(\varphi).
\end{equation}

\item[(b)] Suppose that $f\ge0$ and let $\varphi\in\D_0$ be such that
\begin{equation}
\left\|(W_--1)\e^{-itH_0}\varphi\right\|\in\lone(\R_-,\d t)\label{H-}
\end{equation}
and
\begin{equation}
\left\|(W_+-1)\e^{-itH_0}S\varphi\right\|\in\lone(\R_+,\d t).\label{H+}
\end{equation}
Then
$$
\lim_{r\to\infty}\tau_r^{\rm in}(\varphi)
=\lim_{r\to\infty}\tau_r(\varphi)
=\lim_{r\to\infty}\tau^{\rm free}_r(\varphi).
$$
\end{enumerate}
\end{Lemma}

\begin{proof}
(a) Formula \eqref{evolution} and Assumption \ref{assumption_f}.(i) give
$$
T^0_r(\varphi)=\int_\R\d t\int_\R\d k\,|(\F\varphi)(k)|^2\textstyle f\big(\frac{t-k}r\big).
$$
Then Fubini's theorem (which is applicable due to Assumption \ref{assumption_f}.(ii)) and
the change of variable $x:=\frac{t-k}r$ imply the first identity in \eqref{identities}. The
remaining identities follow from the first identity.

(b) The first equality follows from the third identity in point (a). Since $f\ge0$ one has
\begin{align}
\tau_r(\varphi)-\tau^{\rm free}_r(\varphi)
=&\int_0^\infty\d t\Big[\big\|f(\textstyle\frac Pr)^{1/2}\e^{-itH}W_-\varphi\big\|^2
-\big\|f(\textstyle\frac Pr)^{1/2}\e^{-itH_0}S\varphi\big\|^2\Big]\label{difference}\\
&+\int_{-\infty}^0\d t\Big[\big\|f(\textstyle\frac Pr)^{1/2}\e^{-itH}W_-\varphi\big\|^2
-\big\|f(\textstyle\frac Pr)^{1/2}\e^{-itH_0}\varphi\big\|^2\Big].\nonumber
\end{align}
Using the inequality
$$
\left|\|u\|^2-\|v\|^2\right|\le\|u-v\|\(\|u\|+\|v\|\),\quad u,v\in\H,
$$
and the completeness of $W_\pm$, we obtain the estimates
\begin{align}
\Big|\big\|f(\textstyle\frac Pr)^{1/2}\e^{-itH}W_-\varphi\big\|^2
-\big\|f(\textstyle\frac Pr)^{1/2}\e^{-itH_0}\varphi\big\|^2\Big|
&\le{\rm Const.}\,f_-(t)\,\|\varphi\|\label{borne1}\\
\Big|\big\|f(\textstyle\frac Pr)^{1/2}\e^{-itH}W_-\varphi\big\|^2
-\big\|f(\textstyle\frac Pr)^{1/2}\e^{-itH_0}S\varphi\big\|^2\Big|
&\le{\rm Const.}\,f_+(t)\,\|\varphi\|\label{borne2},
\end{align}
where
$$
f_-(t):=\left\|(W_--1)\e^{-itH_0}\varphi\right\|\quad{\rm and}\quad
f_+(t):=\left\|(W_+-1)\e^{-itH_0}S\varphi\right\|.
$$
We know from Hypotheses \eqref{H-}-\eqref{H+} that $f_\pm\in\lone(\R_\pm,\d t)$. Furthermore
since $\slim_{r\to\infty}f(\frac Pr)^{1/2}=1$, the scalars on the l.h.s. of
\eqref{borne1}-\eqref{borne2} converge to $0$ as $r\to\infty$. Therefore the claim follows from
\eqref{difference} and Lebesgue's dominated convergence theorem.
\end{proof}

\begin{Remark}
The ``velocity" operator associated to the free evolution group $\e^{itQ}$ is constant due to the
canonical commutation rule, namely
$$
\frac\d{\d t}\,(\e^{itQ}P\e^{-itQ})=-1.
$$
Therefore the propagation speed of a state $\e^{itQ}\varphi$ in the space of momenta is equal to
$-1$. In that respect the identities of Lemma \ref{lemma_free}.(a) are natural. For example, if
$\|\varphi\|=1$ and $f=\chi_J$ is as in Example \ref{projection}, then
$T^0_r(\varphi)=r|J|$, where $|J|$ is the Lebesgue measure of $J$. In such a case
$T^0_r(\varphi)$ is nothing else but the sojourn time in $rJ$ (in the space of momenta) of the
state $\e^{itQ}\varphi$ propagating at speed $-1$.
\end{Remark}

Next Theorem is a direct consequence of Formula \eqref{tau_free}, Proposition \ref{equal_P}, and
Lemma \ref{lemma_free}.(b).

\begin{Theorem}\label{abs_delay}
Let $f\ge0$ satisfy Assumption \ref{assumption_f}. Suppose that Assumption \ref{wave} holds. For
some $s>1$, let $\varphi\in\D_s$ satisfy \eqref{H-}-\eqref{H+} and $S\varphi\in\D_s$. Then
\begin{equation}\label{com_P}
\lim_{r\to\infty}\tau^{\rm in}_r(\varphi)=\<\varphi,S^*[P,S]\varphi\>.
\end{equation}
\end{Theorem}

\begin{Remark}
Formula \eqref{com_P} shows that $\lim_{r\to\infty}\tau^{\rm in}_r(\varphi)$ is null if the commutator
$[P,S]$ vanishes (which happens if and only if the scattering operator $S$ is constant). We give an
example of Hamiltonian $H$ for which this occurs.

Let $\widetilde{H_0}:=P$ with domain $\dom(\widetilde{H_0}):=\H^1$, and for $q\in\H$ let
$\widetilde H:=\widetilde{H_0}+q(Q)$ with domain
$\dom(\widetilde H):=\big\{\varphi\in\H^1\mid\widetilde H\varphi\in\H\big\}$. It is known
\cite[Sec. 2.4.3]{Yaf92} that $\widetilde H$ is selfadjoint, that the wave operators
$\widetilde{W_\pm}:=\slim_{s\to\pm\infty}\e^{it\widetilde H}\e^{-it\widetilde{H_0}}$ exist and
are complete, and that $\widetilde S:=\widetilde{W_+}^*\widetilde{W_-}=\e^{-i\int_\R\d x\;\!q(x)}$
is constant. Therefore $H:=\F\widetilde H\F^{-1}=H_0+q(-P)$ is selfadjoint on
$\dom(H):=\F\dom(\widetilde H)$, the wave operators $W_\pm=\F\widetilde{W_\pm}\F^{-1}$ exist and
are complete, and $S=\widetilde S$.
\end{Remark}

\begin{Remark}\label{Remark_Eisenbud}
Suppose that the assumptions of Theorem \ref{abs_delay} are verified, and for a.e. $x\in\R$ let
$S(x)\in\C$ be the component at energy $x$ of the scattering matrix associated to the scattering operator $S$.
Then Equation \eqref{com_P} can be rewritten as
\begin{equation}\label{Eisenbud}
\lim_{r\to\infty}\tau^{\rm in}_r(\varphi)=-i\int_\R\d x\,|\varphi(x)|^2\overline{S(x)}S'(x)
\end{equation}
if the function $x\mapsto S(x)$ is continuously differentiable on the support of $\varphi$
(note that Equation \eqref{Eisenbud} does not follow from \cite{Mar76} or \cite[Chap. 7.2]{AJS77},
since we do not require $f(P/r)$ to be an orthogonal projection or $x\mapsto S(x)$ to be twice
differentiable on the whole real line). Formula \eqref{Eisenbud} holds for the general class of
functions $f\ge0$ satisfying Assumption \ref{assumption_f}. However, if $\|\varphi\|=1$ and $f=\chi_J$
is as in Example \ref{projection}, then we know that the scalars $T^0_r(\varphi)$ and $T_r(\varphi)$
can be interpreted as sojourn times. Therefore in such a case Formula \eqref{Eisenbud} expresses
the identity of the global time delay and the Eisenbud-Wigner time delay for the Friedrichs model.
\end{Remark}

\begin{Remark}\label{RemarkBirman}
Let $R_0(\cdot)$ and $R(\cdot)$ be the resolvent families of $H_0$ and $H$, and suppose that
$R(i)-R_0(i)$ is trace class. Then, at least formally, we get from the Birman-Krein formula
\cite[Thm. 8.7.2]{Yaf92} that
\begin{equation}\label{too_much}
\overline{S(x)}S'(x)=-2\pi i\xi'(x;H,H_0),
\end{equation}
where $\xi'(x;H,H_0)$ is the derivative of the spectral shift function for the pair $\{H_0,H\}$.
Therefore one has
\begin{equation}\label{too_too}
\lim_{r\to\infty}\tau^{\rm in}_r(\varphi)=-2\pi\int_\R\d x\,|\varphi(x)|^2\xi'(x;H,H_0),
\end{equation}
and the number $-2\pi\xi'(x;H,H_0)$ may be interpreted as the component at energy $x$ of the
time delay operator for the Friedrichs model. However Equations \eqref{too_much}-\eqref{too_too}
turn out to be difficult to prove rigorously under this form. We refer to \cite{JSM72},
\cite[Sec. III.b]{Mar81}, and \cite[Sec. 3]{Rob94} for general theories on this issue, and to \cite{Bus71,Oga78,Yaf80} for related works in the case of the Friedrichs-Faddeev model.
\end{Remark}

\section{Finite rank perturbation}\label{finite_rank}

Here we apply the theory of Section \ref{time_delay} to finite rank perturbations of $H_0=Q$. Given
$u,v\in\H$ we write $P_{u,v}$ for the rank one operator $P_{u,v}:=\<u,\;\!\cdot\;\!\>v$, and we set
$P_v:=P_{v,v}$. The full Hamiltonian we consider is defined as follows.

\begin{Assumption}\label{assumption_H}
Fix an integer $N\ge0$ and take $\mu\ge0$. For $j,k\in\{1,\ldots,N\}$, let $v_j\in\H^\mu$ satisfy $\<v_j,v_k\>=\delta_{jk}$, and let $\lambda_j\in\R$. Then $H:=H_0+V$, where
$V:=\sum_{j=1}^N\lambda_jP_{v_j}$.
\end{Assumption}

\noindent
Many functions $v_j$ (as the Hermite functions \cite[p. 142]{RSI}) satisfy the requirements of
Assumption \ref{assumption_H}. Under Assumption \ref{assumption_H} the perturbation $V$ is bounded
from $\H^{-\mu}$ to $\H^\mu$, $H$ is selfadjoint on $\dom(H)=\dom(H_0)$, and the wave operators
$W_\pm$ exist and are complete \cite[Thm. XI.8]{RSIII}.

In the next lemma we establish some of the spectral properties of $H$, we prove a limiting
absorption principle for $H$, and we give a class of locally $H$-smooth operators. The limiting
absorption principle is expressed in terms of the Besov space
$\K:=(\H^1,\H)_{1/2,1}\equiv\H^{1/2,1}$ defined by real interpolation \cite[Chap. 2]{ABG}. We
recall that for each $s>1/2$ we have the continuous embeddings \cite[p. 11]{BdMG96}
$$
\H^s\subset\K\subset\H\subset\K^*\subset\H^{-s}.
$$
We refer the reader to \cite[Sec. 6.2.1]{ABG} for the definition of the regularity classes
$C^k(A)$ and to \cite[Sec. 7.2.2]{ABG} for the definition of a (strict) Mourre estimate. The symbol
$\C_\pm$ stands for the half-plane $\C_\pm:=\{z\in\C\mid\pm\im(z)>0\}$.

\begin{Lemma}\label{rhube}
Let $H$ satisfy Assumption \ref{assumption_H} with $\mu\ge2$. Then
\begin{enumerate}
\item[(a)] $H$ has at most a finite number of eigenvalues, and each of these eigenvalues is of finite
multiplicity.
\item[(b)] The map $z\mapsto(H-z)^{-1}\in\B(\K,\K^*)$, which is holomorphic on $\C_\pm$, extends to
a weak* continuous function on $\C_\pm\cup\{\R\setminus\sigma_{\rm pp}(H)\}$. In particular, $H$ has
no singularly continuous spectrum.
\item[(c)] If $T$ belongs to $\B(\H^{-s},\H)$ for some $s>1/2$, then $T$ is locally $H$-smooth on
$\R\setminus\sigma_{\rm pp}(H)$.
\end{enumerate}
\end{Lemma}

The spectral results of points (a) and (b) on the finiteness of the singular spectrum of $H$ are not
surprising; they are known in the more general setting where $V$ is an integral operator with
H\"older continuous kernel (see \eg \cite[Thm. 1]{DNY91} and \cite[Lemma 3.10]{Fad64}). Note
however that point (a) implies that the sets $\D_s$ are dense in $\H$ for each $s\ge0$.

\begin{proof}
(a) Let $A:=-P$, then $\e^{-itA}H_0\e^{itA}=H_0+t$ for each $t\in\R$. Thus $H_0$ is of class
$C^\infty(A)$ and satisfies a strict Mourre estimate on $\R$ \cite[Sec. 7.6.1]{ABG}. Furthermore the
quadratic form
$$
\dom(A)\ni\varphi\mapsto\<A\varphi,iV\varphi\>-\<\varphi,iVA\varphi\>
$$
extends uniquely to the bounded form defined by the rank $2N$ operator
$F_1:=-\sum_{j=1}^N\lambda_j\big(P_{v_j,v_j'}+P_{v_j',v_j}\big)$. This means that $V$ is of class
$C^1(A)$. Thus $H$ is of class $C^1(A)$ and since $F_1$ is compact, $H$ satisfies a Mourre estimate
on $\R$. The claim then follows by \cite[Cor. 7.2.11]{ABG}.

(b) The quadratic form
$$
\dom(A)\ni\varphi\mapsto\<A\varphi,iF_1\varphi\>-\<\varphi,iF_1A\varphi\>
$$
extends uniquely to the bounded form defined by the rank $3N$ operator
$F_2:=\sum_{j=1}^N\lambda_j\big(P_{v_j'',v_j}+2P_{v_j',v_j'}+P_{v_j,v_j''}\big)$. This, together with
\cite[Thm. 7.2.9 \& Thm. 7.2.13]{ABG} and the proof of point (a), implies that $H$ is of class
$C^2(A)$ and that $H$ satisfies a strict Mourre estimate on $\R\setminus\sigma_{\rm pp}(H)$. It follows by
\cite[Thm. 01]{Sah97} (which applies to operators without spectral gap) that the map
$z\mapsto(H-z)^{-1}\in\B(\K,\K^*)$ extends to a weak* continuous function on
$\C_\pm\cup\{\R\setminus\sigma_{\rm pp}(H)\}$. In particular, $H$ has no singularly continuous
spectrum in $\R\setminus\sigma_{\rm pp}(H)$. Since continuous Borel measures on $\R$ have no pure
points \cite[p. 22]{RSI} and since $\sigma_{\rm pp}(H)$ is finite by point (a), we even get that $H$
has no singularly continuous spectrum at all.

(c) Since $T$ belongs to $\B(\dom(H),\H)$ and $T^*\H\subset\H^s\subset\K$, the claim is a consequence
of \cite[Prop. 7.1.3.(b)]{ABG} and the discussion that follows.
\end{proof}

We now study the differentiability of the function $x\mapsto S(x)$, which relies on the
differentiability of the boundary values of the resolvent of $H$.

\begin{Lemma}\label{higher_order}
Let $H$ satisfy Assumption \ref{assumption_H} with $\mu\ge n+1$ for some integer $n\ge1$. Let
$I\subset\{\R\setminus\sigma_{\rm pp}(H)\}$ be a relatively compact interval, and take $s>n-1/2$.
Then for each $x\in I$ the limits
$$
R^n(x\pm i0):=\lim_{\varepsilon\searrow0}(H-x\mp i\varepsilon)^{-n}
$$
exist in the norm topology of $\B(\H^s,\H^{-s})$ and are H\"older continuous. Furthermore
$x\mapsto R(x\pm i0)$ is $n-1$ times (H\"older continuously) differentiable as a map from $I$
to $\B(\H^s,\H^{-s})$, and
$$
\frac{\d^{n-1}}{\d x^{n-1}}\;\!R(x\pm i0)=(n-1)!\;\!R^n(x\pm i0).
$$
\end{Lemma}

\begin{proof}
The claims follow from \cite[Thm. 2.2.(iii)]{JMP84} applied to our
situation. We only have to verify the hypotheses of that theorem, namely that $H$ is $n$-smooth
with respect to $A=-P$ in the sense of \cite[Def. 2.1]{JMP84}. This is done in points (a), (b),
($\rm c_n$), ($\rm d_n$), and (e) that follow.

(a) $\dom(A)\cap\dom(H)\supset\S$ is dense in $\dom(H)$.

(b) Let $\varphi\in\H_1$ and $\theta\in\R$. Then one has
$$
\|\e^{i\theta A}\varphi\|_{\H_1}
=\|\<Q+\theta\>\varphi\|
\le\big\|\<Q+\theta\>\<Q\>^{-1}\big\|\cdot\|\varphi\|_{\H_1}
\le2^{-1/2}(2+|\theta|)^{1/2}\|\varphi\|_{\H_1}.
$$
In particular, $\e^{i\theta A}$ maps $\dom(H)$ into $\dom(H)$, and
$\sup_{|\theta|\le1}\|H\e^{i\theta A}\varphi\|<\infty$ for each $\varphi\in\dom(H)$.

($\rm c_n$)-($\rm d_n$) Due to the proof Lemma \ref{rhube}.(a) the quadratic form
$$
\dom(A)\cap\dom(H)\ni\varphi\mapsto\<A\varphi,iH\varphi\>-\<H\varphi,iA\varphi\>
$$
extends uniquely to the bounded form defined by the operator $iB_1:=1+F_1$, where
$F_1=-\sum_{j=1}^N\lambda_j\big(P_{v_j',v_j}+P_{v_j,v_j'}\big)$. Similarly for $j=2,3,\ldots,n+1$
the quadratic form
$$
\dom(A)\cap\dom(H)\ni\varphi\mapsto\<A\varphi,i(iB_{j-1})\varphi\>
-\<(iB_{j-1})^*\varphi,iA\varphi\>
$$
extends uniquely to a bounded form defined by an operator $iB_j:=F_j$, where $F_j$ is a
linear combination of the rank one operators $P_{v^{(j-k)},v^{(k)}}$, $k=0,1,\ldots,j$.

(e) Due to the proof Lemma \ref{rhube}.(a), $H$ satisfies a Mourre estimate on $\R$.
\end{proof}

For $m=1,2,\ldots,N$ let $V_m:=\sum_{j=1}^m\lambda_jP_{v_j}$ and $H_m:=H_0+V_m$. Then
it is known that the scattering matrix $S(x)$ factorizes for a.e. $x\in\R$ as
\cite[Eq. (8.4.2)]{Yaf92}
\begin{equation}\label{S_product}
S(x)=\widetilde{S_N}(x)\cdots\widetilde{S_2}(x)\widetilde{S_1}(x),
\end{equation}
where $\widetilde{S_m}(x)$ is unitarily equivalent to the scattering matrix $S_m(x)$
associated to the pair $\{H_m,H_{m-1}\}$. Since the difference $H_m- H_{m-1}$ is of rank one,
one can even obtain an explicit expression for $S_m(x)$ (see \cite[Eq. (6.7.9)]{Yaf92}).
For instance one has the following simple formula for $S_1(x)$ \cite[Eq. (8.4.1)]{Yaf92},
\cite[Eq. (66a)]{GP70}
$$
S_1(x)=\frac{1+\lambda_1F(x-i0)}{1+\lambda_1F(x+i0)}\;\!,
$$
where
$$
F(x\pm i0):=\lim_{\varepsilon\searrow0}\<v_1,(H_0-x\mp i\varepsilon)^{-1}v_1\>.
$$
Clearly Formula \eqref{S_product} is not very convenient for studying the differentiability of
the function $x\mapsto S(x)$. This is why we prove the usual formula for $S(x)$ in the next lemma.

Given $\tau\in\R$, we
let $\gamma(\tau):\S\to\C$ be the restriction operator defined by $\gamma(\tau)\varphi:=\varphi(\tau)$.
Some of the regularity properties of $\gamma(\tau)$ are collected in the appendix. Here we only recall that
$\gamma(\tau)$ extends uniquely to an element of $\B(\H^s,\C)$ for each $s>1/2$.

\begin{Lemma}\label{representation}
Let $H$ satisfy Assumption \ref{assumption_H} with $\mu\ge2$. Then for each
$x\in\R\setminus\sigma_{\rm pp}(H)$ one has the equality
\begin{equation}\label{S_formula}
S(x)=1-2\pi i\gamma(x)[1-VR(x+i0)]V\gamma(x)^*.
\end{equation}
\end{Lemma}
\begin{proof}
The claim is a consequence of the stationary method for trace class perturbations \cite[Thm. 7.6.4]{Yaf92}
applied to the pair $\{H_0,H\}$.

The perturbation $V$ can be written as a product $V=G^*G_0$, with $G:=\sum_{j=1}^N\lambda_jP_{v_j}$ and
$G_0:=\sum_{j=1}^NP_{v_j}$. Since the operators $G$ and $G_0$ are selfadjoint and belong to the
Hilbert-Schmidt class, all the hypotheses of \cite[Thm. 7.6.4]{Yaf92} (and thus of
\cite[Thm. 5.7.1]{Yaf92}) are trivially satisfied. Therefore one has for a.e. $x\in\R$ the equality
\begin{equation}\label{Yafaev}
S(x)=1-2\pi i\gamma(x)G\big[1-\widetilde B(x+i0)\big]G_0\gamma(x)^*,
\end{equation}
where $\widetilde B(x+i0)$ is the norm limit defined by the condition
$$
\lim_{\varepsilon\searrow0}\big\|G_0(H-x-i\varepsilon)^{-1}G-\widetilde B(x+i0)\big\|=0.
$$
On another hand we know from Lemma \ref{higher_order} that the limit $R(x+i0)$ exists in the
norm topology of $\B(\H^s,\H^{-s})$ for each $x\in\R\setminus\sigma_{\rm pp}(H)$ and each
$s>1/2$. Since we also have $G_0,G\in\B(\H^{-\mu},\H^\mu)$, we get the identity
$\widetilde B(x+i0)=G_0R(x+i0)G$. This together with Formula \eqref{Yafaev} implies the claim.
\end{proof}

We are in a position to show the differentiability of the scattering matrix.

\begin{Lemma}\label{S_differentiable}
Let $H$ satisfy Assumption \ref{assumption_H} with $\mu\ge n+1$ for some integer $n\ge1$. Then
$x\mapsto S(x)$ is $n-1$ times (H\"older continuously) differentiable from $\R\setminus\sigma_{\rm pp}(H)$
to $\C$.
\end{Lemma}

\begin{proof}
Due to Formula \eqref{S_formula} it is sufficient to prove that the terms
$$
A(x):=\big(\textstyle\frac{\d^{\ell_1}}{\d x^{\ell_1}}\;\!\gamma(x)\big)V
\big(\textstyle\frac{\d^{\ell_2}}{\d x^{\ell_2}}\;\!\gamma(x)^*\big)
$$
and
$$
B(x):=\big(\textstyle\frac{\d^{\ell_1}}{\d x^{\ell_1}}\;\!\gamma(x)\big)V
\big(\textstyle\frac{\d^{\ell_2}}{\d x^{\ell_2}}\;\!R(x+i0)\big)V
\big(\textstyle\frac{\d^{\ell_3}}{\d x^{\ell_3}}\;\!\gamma(x)^*\big)
$$
exist and are locally H\"older continuous on $\R\setminus\sigma_{\rm pp}(H)$ for all non-negative
integers $\ell_1,\ell_2,\ell_3$ satisfying $\ell_1+\ell_2+\ell_3\le n-1$. The factors in $B(x)$
satisfy
\begin{align*}
\big(\textstyle\frac{\d^{\ell_3}}{\d x^{\ell_3}}\;\!\gamma(x)^*\big)
&\in\B(\C,\H^{-s_3})\quad{\rm for}~s_3>\ell_3+1/2,\\
V&\in\B(\H^{-s_3},\H^{s_2})\quad{\rm for}~s_2,s_3\in[0,\mu],\\
\big(\textstyle\frac{\d^{\ell_2}}{\d x^{\ell_2}}\;\!R(x+i0)\big)
&\in\B(\H^{s_2},\H^{-s_2})\quad{\rm for}~s_2>\ell_2+1/2,\\
V&\in\B(\H^{-s_2},\H^{s_1})\quad{\rm for}~s_1,s_2\in[0,\mu],\\
\big(\textstyle\frac{\d^{\ell_1}}{\d x^{\ell_1}}\;\!\gamma(x)\big)
&\in\B(\H^{s_1},\C)\quad{\rm for}~s_1>\ell_1+1/2,
\end{align*}
and are locally H\"older continuous due to Lemma \ref{higher_order} and Lemma \ref{gamma_smooth}.
Therefore if the $s_j$'s above are chosen so that $s_j\in(\ell_j+1/2,\mu]$ for $j=1,2,3$, then
$B(x)$ is finite and locally H\"older continuous on $\R\setminus\sigma_{\rm pp}(H)$. Since similar
arguments apply to the term $A(x)$, the claim is proved.
\end{proof}

\begin{Lemma}\label{lone_conditions}
Let $H$ satisfy Assumption \ref{assumption_H} with $\mu>2$. Then one has for each $\varphi\in\D_s$,
$s>2$,
\begin{equation}
\left\|(W_--1)\e^{-itH_0}\varphi\right\|\in\lone(\R_-,\d t)\label{R-}
\end{equation}
and
\begin{equation}
\left\|(W_+-1)\e^{-itH_0}\varphi\right\|\in\lone(\R_+,\d t).\label{R+}
\end{equation}
\end{Lemma}

\begin{proof}
For $\varphi\in\D_s$ and $t\in\R$, we have (see \eg the proof of \cite[Lemma 4.6]{Jen81})
$$
\(W_--1\)\e^{-itH_0}\varphi
=-i\e^{-itH}\int_{-\infty}^t\d\tau\,\e^{i\tau H}V\e^{-i\tau H_0}\varphi,
$$
where the integral is strongly convergent. Hence to prove \eqref{R-} it is enough to show that
\begin{equation}\label{l1 condition}
\int_{-\infty}^{-\delta}\d t\int_{-\infty}^t\d\tau\left\|V\e^{-i\tau H_0}\varphi\right\|<\infty
\end{equation}
for some $\delta>0$. Let $\zeta:=\min\{\mu,s\}$, then $\big\|\<P\>^\zeta\varphi\big\|$ and
$\big\|V\<P\>^\zeta\big\|$ are finite by hypothesis. If $|\tau|$ is big enough, it follows that
\begin{align*}
\big\|V\e^{-i\tau H_0}\varphi\big\|
\le{\rm Const.}\;\!\big\|\<P\>^{-\zeta}\e^{-i\tau Q}\<P\>^{-\zeta}\big\|
&={\rm Const.}\;\!\big\|\<P-\tau\>^{-\zeta}\<P\>^{-\zeta}\big\|\\
&\le{\rm Const.}\;\!|\tau|^{-\zeta}.
\end{align*}
Since $\zeta>2$, this implies \eqref{l1 condition}, and thus \eqref{R-}. The proof of \eqref{R+}
is similar.
\end{proof}

In the next theorem we prove the existence of time delay and its identity with Eisenbud-Wigner
time delay for Hamiltonians $H$ satisfying Assumption \ref{assumption_H} with $\mu\ge5$.

\begin{Theorem}\label{final_one}
Let $f\ge0$ satisfy Assumption \ref{assumption_f}, and let $H$ satisfy Assumption
\ref{assumption_H} with $\mu\ge5$. Then one has for each $\varphi\in\D_3$ the identity
$$
\lim_{r\to\infty}\tau^{\rm in}_r(\varphi)=-i\int_\R\d x\,|\varphi(x)|^2\overline{S(x)}S'(x).
$$
\end{Theorem}

\begin{proof}
Let $\varphi\in\D_3$. Then $S\varphi\in\D_3$ by Lemma \ref{S_differentiable}, and conditions
\eqref{H-}-\eqref{H+} are verified by Lemma \ref{lone_conditions}. Therefore all the hypotheses of
Theorem \ref{abs_delay} and Remark \ref{Remark_Eisenbud} are satisfied, and so the claim is
proved.
\end{proof}

\section*{Acknowledgements}

The author thanks the Swiss National Science Foundation for financial support.

\section*{Appendix}

We collect in this appendix some facts on the restriction operator $\gamma(\tau)$ of Lemma
\ref{representation}. We consider the general case with configurations space $\R^d$, $d\ge1$.

Let $P\equiv(P_1,P_2,\ldots,P_d)$ be the vector momentum operator in $\ltwo(\R^d)$. For each
$s\in\R$, we denote by $\H^s(\R^d)$ the completion of the Schwartz space $\S(\R^d)$ on $\R^d$ for
the norm $\|\varphi\|_{\H^s(\R^d)}:=\|\<P\>^s\varphi\|$. Given $\tau\in\R$, we let $\gamma(\tau):\S(\R^d)\to\ltwo(\R^{d-1})$ be the restriction operator defined by $\gamma(\tau)\varphi:=\varphi(\tau,\cdot)$. We know from \cite[Thm. 2.4.2]{Kur78} that
$\gamma(\tau)$ extends uniquely to an element of $\B(\H^s(\R^d),\ltwo(\R^{d-1}))$ for each $s>1/2$.
Furthermore $\gamma(\tau)$ is H\"older continuous in $\tau$ with respect to the operator norm,
namely for each $\tau,\tau'\in\R$ there exists a constant $\textsc c$ such that
\begin{equation}\label{Holder}
\big\|\gamma(\tau)-\gamma(\tau')\big\|_{\B(\H^s(\R^d),\ltwo(\R^{d-1}))}\le\textsc c
\begin{cases}
|\tau-\tau'|^{s-1/2} & \textrm{if }s\in\big(\12,\frac32\big),\\
|\tau-\tau'|\cdot|\ln|\tau-\tau'|| & \textrm{if }s=\frac32\textrm{ and }|\tau-\tau'|<\12,\\
|\tau-\tau'| & \textrm{if }s>\frac32.
\end{cases}
\end{equation}
Finally $\gamma(\tau)$ has the following differentiability property.

\begin{Lemma}\label{gamma_smooth}
Let $s>k+\12$ with $k\ge0$ integer. Then $\gamma$ is $k$ times (H\"older continuously) differentiable as
a map from $\R$ to $\B(\H^s(\R^d),\ltwo(\R^{d-1}))$.
\end{Lemma}

\begin{proof}
We adapt the proof of \cite[Lemma 3.3]{Jen81}. Consider first $s>k+\12$ with $k=1$. The
obvious guess for the derivative at $\tau$ of $\gamma$ is
$({\sf D}\gamma)(\tau):=\gamma(\tau)\partial_1$, where $\partial_1$ stands for the partial
derivative w.r.t. the first variable. Thus one has for $\varphi\in\S(\R^d)$ and
$\delta\in\R$ with $|\delta|\in(0,1/2)$
$$
\big\{{\textstyle\frac1\delta}[\gamma(\tau+\delta)-\gamma(\tau)]-({\sf D}\gamma)(\tau)\big\}\varphi
={\textstyle\frac1\delta}\int_0^\delta\d\xi
\big[(\partial_1\varphi)(\tau+\xi,\;\!\cdot\;\!)-(\partial_1\varphi)(\tau,\;\!\cdot\;\!)\big].
$$
In particular, using the first (and thus the most pessimistic) bound in \eqref{Holder}, we get
\begin{align*}
&\big\|\big\{{\textstyle\frac1\delta}[\gamma(\tau+\delta)
-\gamma(\tau)]-({\sf D}\gamma)(\tau)\big\}\varphi\big\|_{\ltwo(\R^{d-1})}\\
&\le{\textstyle\frac1{|\delta|}}\int_0^{|\delta|}\d\xi\;\!
\big\|(\partial_1\varphi)(\tau+\sgn(\delta)\xi,\;\!\cdot\;\!)
-(\partial_1\varphi)(\tau,\;\!\cdot\;\!)\big\|_{\ltwo(\R^{d-1})}\\
&\le\|\partial_1\varphi\|_{\H^{s-1}(\R^d)}\;\!{\textstyle\frac1{|\delta|}}\int_0^{|\delta|}\d\xi\,
\|\gamma(\tau+\sgn(\delta)\xi)-\gamma(\tau)\|_{\B(\H^{s-1}(\R^d),\ltwo(\R^{d-1}))}\\
&\le{\rm Const.}\,\|\varphi\|_{\H^s(\R^d)}\;\!{\textstyle\frac1{|\delta|}}\int_0^{|\delta|}\d\xi\,
|\xi|^{s-3/2}\\
&\le{\rm Const.}\,\|\varphi\|_{\H^s(\R^d)}|\delta|^{s-3/2}.
\end{align*}
Since $\S(\R^d)$ is dense in $\H^s(\R^d)$ and ${\sf D}\gamma:\R\to\B(\H^s(\R^d),\ltwo(\R^{d-1}))$ is H\"older continuous, the result is proved for $k=1$. The result for $k>1$ follows then easily by
using the expression for $({\sf D}\gamma)(\tau)$.
\end{proof}


\def\cprime{$'$}

\end{document}